\documentclass[sigconf]{acmart}

\usepackage{amsmath}
\usepackage{booktabs}
\usepackage{array}
\usepackage{graphicx}

\AtBeginDocument{%
  }

\setcopyright{none}
\copyrightyear{2027}
\acmYear{2027}
\acmDOI{}
\acmConference[KDD '27]{Proceedings of the 33rd ACM SIGKDD Conference on Knowledge Discovery and Data Mining}{August 2027}{San Jose, CA, USA}
\acmISBN{}
\renewcommand\footnotetextcopyrightpermission[1]{}

\newcommand{\systemname}{MetaStrategy}
\newcommand{\guly}{Guess You Like}
\newcommand{\toolprocess}{\mathcal{T}}

\makeatletter
\renewcommand{\@authorfont}{\large}
\renewcommand{\@affiliationfont}{\footnotesize\small}
\patchcmd{\maketitle}
  {\noindent\@concepts\par}
  {\noindent\@concepts\par\vfill\newpage}
  {}
  {\PackageWarning{metastrategy}{Could not adjust first-page CCS spacing}}
\makeatother

\begin{document}

\title{\systemname: Generative Ranking with Executable LLM Strategies}

\author{Chengyu Lai}
\author{Jiuning Lin}
\author{Zhibo Xiao}
\affiliation{%
  \institution{Taobao \& Tmall Group of Alibaba}
  \city{Hangzhou}
  \country{China}}
\email{lcy477544@alibaba-inc.com}
\email{linjiuning.ljn@alibaba-inc.com}
\email{xiaozhibo.xzb@alibaba-inc.com}

\author{Xiaodong Zhu}
\affiliation{%
  \institution{Wuhan University}
  \city{Wuhan}
  \country{China}}
\email{xiaodongzhu@whu.edu.cn}

\author{Ruiquan Lan}
\affiliation{%
  \institution{Taobao \& Tmall Group of Alibaba}
  \city{Hangzhou}
  \country{China}}
\email{lanruiquan.lrq@alibaba-inc.com}

\author{Bin Zhang}
\author{Zihong Huang}
\author{Wendong Zhang}
\affiliation{%
  \institution{Taobao \& Tmall Group of Alibaba}
  \city{Hangzhou}
  \country{China}}
\email{tianji.zb@alibaba-inc.com}
\email{huangzihong.hzh@alibaba-inc.com}
\email{xianzhi.zwd@alibaba-inc.com}

\author{Chuxin Chen}
\author{Yinjiang Cai}
\affiliation{%
  \institution{Taobao \& Tmall Group of Alibaba}
  \city{Hangzhou}
  \country{China}}
\email{chenchuxin.ccx@alibaba-inc.com}
\email{caiyinjiang.cyj@alibaba-inc.com}

\author{Shuai Zhong}
\affiliation{%
  \institution{The University of Hong Kong }
  \city{Hong Kong}
  \country{Hong Kong SAR}
  }
\email{chris\_zsa@connect.hku.hk}

\author{Lingqing Zhang}
\affiliation{%
  \institution{University of Cambridge}
  \city{Cambridge}
  \country{United Kingdom}}
\email{lingqing.zhang@mail.mcgill.ca}

\author{Dimin Wang}
\author{Jialin Zhu}
\author{Han Zhu}
\affiliation{%
  \institution{Taobao \& Tmall Group of Alibaba}
  \city{Hangzhou}
  \country{China}}
\email{dimin.wdm@alibaba-inc.com}
\email{xiafei.zjl@alibaba-inc.com}
\email{zhuhan.zh@alibaba-inc.com}

\renewcommand{\shortauthors}{Chengyu Lai et al.}

\begin{abstract}
Industrial recommender systems rank heterogeneous content under coupled user,
business, commercial, and experience objectives. Existing generative ranking
methods typically construct item sequences directly, making them difficult to
integrate with mature predictive models, operational rules, and field-level
guardrails. We present \systemname, a framework that instead generates a
structured, executable ranking strategy. Conditioned on request context, a
large language model (LLM) policy emits a typed JSON bundle controlling
objective weights, content and category preferences, experience constraints,
and position policies. A deterministic validator and compiler instantiate an
isolated Generator that competes atomically with incumbents under the list-level
Evaluator of the Generator-Evaluator (GE) architecture. We train the policy in a
production-path replay environment that re-executes logged requests through the
current re-ranking stack without user exposure. The method combines selection,
relative-rank, and baseline-lift rewards, a self-competitive curriculum that
feeds frequent strategies back as competitors, and Evaluator-routed
reward-augmented on-policy distillation that transfers complementary
4B-parameter Teachers into a compact 0.8B-parameter Student. We deploy
\systemname{} in Taobao Homepage \guly{} through diff-triggered nearline
generation; LLM inference remains outside synchronous ranking, with no
observable increase in response time (RT). In a seven-day user-randomized online
A/B test, \systemname{} wins 27.93\% of treatment-side GE calls and
significantly improves click page views (click PV) by 2.11\%,
item-detail page views (IPV) by 3.12\%, and transaction amount by 2.83\%.
\end{abstract}




\maketitle

\section{Introduction}

Industrial recommendation lists are rarely determined by a single relevance
score. An e-commerce homepage may mix products, content posts, short videos,
live streams, and sponsored items \cite{xu2023multi}. These content types have different
prediction targets and operational constraints, yet they compete for the same
exposure positions. Production systems therefore place a strategy layer above
their predictive models to coordinate objective weights, content composition,
category preferences, exposure filters, density constraints, and
position-specific rules \cite{xu2023multifactor}.

This strategy layer is reliable but difficult to personalize. Global
multi-objective formulas cannot express every request-specific trade-off, and
tuning rules independently ignores their interactions \cite{liu2022neural}. For example, a
click-oriented weight change may be overridden by a top-position policy, while
a content-type boost may violate a density constraint after re-ranking. The
decision must therefore be made jointly, conditioned on user context, current
intent, and available candidates.

Generative recommendation provides a natural way to model such joint
decisions \cite{zhou2025onerec,ren2024non}. Existing work formulates recommendation as text generation,
semantic-identifier generation, or autoregressive slate construction
\cite{geng2022p5,cui2022m6rec,rajput2023tiger}. Directly generating item
sequences, however, creates a difficult integration boundary in a mature
high-traffic system. The generated sequence must reproduce established
filters, calibrated predictors, quotas, and safety controls, while offering no
simple way to inspect or correct an individual business decision.

We instead generate the ranking \emph{strategy} and let the production ranker
execute it. The strategy policy consumes a real-time multi-intent representation
produced by an upstream intent model, historical user statistics, and an ordered
sequence of recent interactions. Given this context, a large language model
(LLM) produces one
schema-constrained JSON bundle that jointly adjusts several ranking modules. A
deterministic tool process validates the bundle and translates its semantic
actions into existing production parameters. The LLM does not emit item
identifiers or an item permutation. This separation retains the expressive
joint policy of an autoregressive model while preserving the control surface of
the deployed ranker.

\systemname{} operates within a Generator-Evaluator (GE)
architecture~\cite{feng2021grn}. Each
Generator transforms a shared candidate set containing hundreds of items into
a list containing tens of items. A discriminative list-level Evaluator scores
the complete lists with user, item, position, and business context, then
selects the list to expose. The LLM controls one isolated Generator and
competes against approximately ten incumbent production Generators. It can
therefore improve the final decision without bypassing the existing production boundary.

Reinforcement learning (RL) in this setting presents two challenges. The
Evaluator is useful but
imperfect, so direct reward optimization can collapse to a few extreme bundles
that exploit stable proxy-model preferences rather than adapt to each request.
Moreover, large LLMs are valuable teachers during offline exploration but are
too expensive for the intended serving budget. We address collapse with a
self-competitive curriculum that freezes frequent compiled strategies as new
Generators in the next training round. We address serving cost with
Evaluator-routed reward-augmented on-policy distillation (OPD): multiple
teachers are compared atomically on each replayed request, and the best valid
teacher supplies an on-policy token signal to a compact student without
replacing the student's GE reward.

We instantiate the framework in Taobao Homepage \guly{}. Strategy generation
runs nearline and is refreshed only when decision-relevant context changes;
the online ranker reads the latest validated strategy and applies it to the
LLM-controlled Generator. The deployed policy is the final 0.8B Student trained
with Evaluator-routed reward-augmented OPD. The online request never waits for
LLM inference, and the lightweight lookup and deterministic resolution add no
observable response-time (RT) overhead. Integration is additive: incumbent Generators, the
list-level Evaluator, and production fallback behavior remain unchanged. In an
online A/B test,
\systemname{} increases exposure page views (exposure PV) by 1.49\%, click page
views (click PV) by 2.11\%, item-detail page views (IPV) by 3.12\%, and
transaction amount by 2.83\% relative to the production control.

Our contributions are:
\begin{itemize}
  \item We formulate generative ranking as request-conditioned generation of
  an executable joint strategy. A typed action schema and deterministic
  compiler let the LLM control multiple production ranking modules while
  preserving validation, isolation, and fallback behavior.
  \item We build a production-path replay environment that compares incumbent
  and LLM-controlled Generators in one call to the current re-ranking stack.
  Selection, relative-rank, and baseline-lift signals provide
  request-level feedback without exposing replay traffic to users.
  \item We introduce a self-competitive curriculum that turns frequent
  compiled strategies into explicit competitors, and an Evaluator-routed,
  reward-augmented OPD objective that compresses complementary teachers while
  retaining direct GE optimization.
  \item We deploy \systemname{} in Taobao Homepage \guly{} through a
  non-intrusive serving architecture: nearline LLM inference adds no observable
  online RT overhead, while an isolated Generator preserves the incumbent
  ranking and fallback paths. We report significant online gains in click PV,
  IPV, and transaction amount.
\end{itemize}

\section{Related Work}
\label{sec:related}

\textbf{Industrial multi-objective ranking.}
Large-scale recommenders commonly separate candidate generation and ranking
\cite{covington2016youtube}, while multi-task architectures such as the Entire
Space Multi-Task Model (ESMM) and Multi-gate Mixture-of-Experts (MMoE) improve
estimates for coupled engagement and transaction objectives
\cite{ma2018esmm,ma2018mmoe}. Industrial mixed-content feeds must additionally
coordinate cross-channel exposure constraints and perception-aware
diversification \cite{xu2023multi,xu2023multifactor}. Personalized re-ranking
also studies popularity and diversity trade-offs
\cite{abdollahpouri2019popularity,kunaver2017diversity}. These approaches
improve predictive signals or a particular re-ranking objective.
\systemname{} instead learns a request-level policy that jointly configures
multiple objectives and operational rules while leaving the predictive stack
intact.

\textbf{Listwise re-ranking and slate optimization.}
Seq2Slate autoregressively constructs a slate, and SLATEQ develops tractable
reinforcement learning over slate actions
\cite{bello2018seq2slate,ie2019slateq}. Personalized re-rankers model
interactions among candidates in the final list
\cite{pei2019personalized,li2022pear}. Generative re-rankers further model
complete-list construction through context-wise, non-autoregressive, or
permutation-level objectives \cite{feng2021grn,ren2024non,xu2026omgrec}.
Their principal action is still an item order or permutation. Our action is a
bounded executable strategy; list construction remains delegated to the
production ranker, and the resulting list is selected under the incumbent
list-level Evaluator.

\textbf{Generative and LLM-based recommendation.}
P5 and M6-Rec formulate recommendation tasks as language modeling
\cite{geng2022p5,cui2022m6rec}; LLMRank applies an LLM as a zero-shot ranker
\cite{hou2023llmrank}; and TIGER and OneRec generate semantic identifiers or
recommendation sequences \cite{rajput2023tiger,zhou2025onerec}. Surveys
summarize the rapidly expanding design space of LLM-based and generative
recommendation \cite{wu2023llmsurvey,li2023generativesurvey}. In contrast,
\systemname{} does not ask the LLM to reproduce retrieval, scoring, filtering,
or list construction. It uses the LLM as a typed strategy policy over existing
production mechanisms, preserving their validation, guardrails, and fallback
paths.

\textbf{Reinforcement learning and distillation.}
Recommendation and ranking have been optimized with reinforcement learning
over item or slate actions \cite{hu2018rlrank,ie2019slateq}. Conventional
knowledge distillation transfers a fixed Teacher to a smaller Student
\cite{hinton2015distilling}, while on-policy distillation evaluates Teacher
feedback on Student-generated trajectories to reduce distribution mismatch
\cite{agarwal2024onpolicy}. Our learning signal instead evaluates an executed
strategy against production Generators. Frequent compiled modes become new
competitors in the curriculum, and the distillation Teacher is routed per
request through a shared list-level evaluation before its token signal is
combined with the Student's GE reward.

\begin{figure*}[t]
  \centering
  \includegraphics[width=0.8\textwidth]{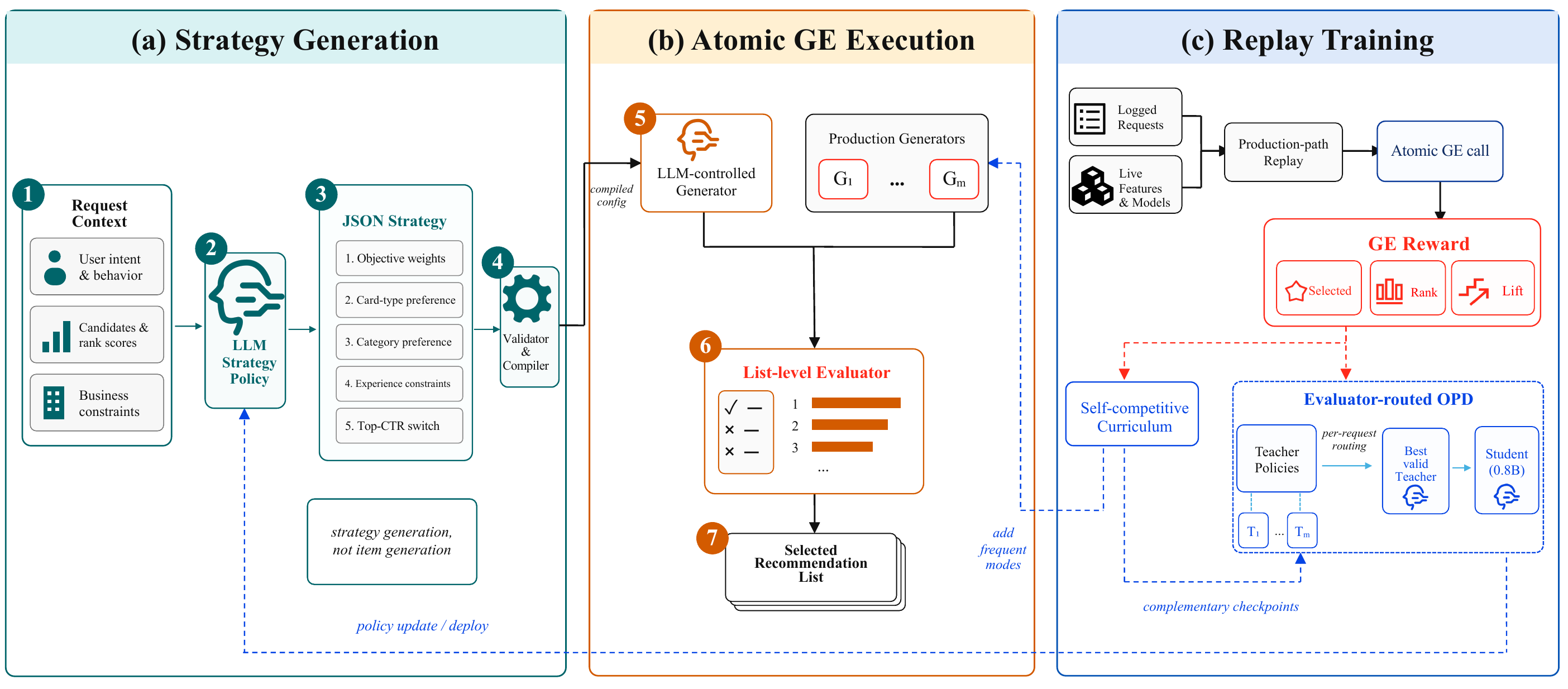}
  \caption{Overview of \systemname{}. (a) The LLM maps request context to one
  structured JSON strategy, which is validated and compiled rather than
  directly generating an item sequence. (b) The compiled configuration
  controls one Generator that competes with production Generators under a
  shared list-level Evaluator. (c) Production-path replay supplies the three GE
  rewards used for self-competitive curriculum learning and per-request
  evaluator-routed OPD into the deployed 0.8B Student.}
  \Description{A three-panel workflow. Panel a shows request context, including
  user intent and behavior, candidates and rank scores, and business
  constraints, flowing through an LLM strategy policy to a five-module JSON
  strategy and then a validator and compiler. Panel b shows the resulting
  LLM-controlled Generator and production Generators entering one list-level
  Evaluator, which selects a recommendation list. Panel c shows logged requests
  and live production features and models entering production-path replay and
  an atomic GE call. Selection, rank, and lift rewards drive a
  self-competitive curriculum and evaluator-routed on-policy distillation.
  Dashed feedback paths add frequent modes to the competing Generator pool,
  contribute complementary checkpoints to the Teacher pool, and update the
  deployed 0.8B Student policy.}
  \label{fig:overview}
\end{figure*}

\section{Setting and Problem Formulation}
\label{sec:problem}

\subsection{Industrial Strategy Composition}

Let $x$ denote the context of one recommendation request and
$I=\{i_1,\ldots,i_n\}$ its shared candidate set. The user-side context contains
three complementary views: (1) a real-time multi-intent representation from an
upstream intent model; (2) historical statistics that summarize longer-term
preferences and engagement; and (3) an ordered real-time behavior sequence,
including clicks, add-to-cart events, purchases, and other recent interactions.
It is combined with request state and the candidate-supply summary required for
strategy decisions. In our deployment, $n$ is on the order of hundreds, while
the exposed list contains tens of items.

A production ranking strategy is a composition of decisions rather than one
scalar score. A simplified objective fusion can be written as
\begin{equation}
  q(i\mid x)=
  \alpha(x)\,\widehat{v}_{\mathrm{eng}}(i,x)
  +\beta(x)\,\widehat{v}_{\mathrm{txn}}(i,x)
  +\gamma(x)\,\widehat{v}_{\mathrm{com}}(i,x),
\end{equation}
but the final list also depends on filtering, supply mixing, diversity, and
position rules. Existing multi-task models improve the underlying predictions
\cite{ma2018esmm,ma2018mmoe}; our focus is the request-level policy that
coordinates these predictions and rules.
We use CTR, CVR, and GMV for click-through rate, conversion rate, and gross
merchandise value, respectively; the prefix \(p\) denotes a model prediction.

\subsection{Generator-Evaluator Architecture}

For strategy $s_j$, Generator $G_j$ maps the shared candidates to an ordered
list:
\begin{equation}
  L_j = G_j(I,x;s_j), \qquad |L_j| \ll |I|.
  \label{eq:generator}
\end{equation}
The Generators may represent existing ranking recipes, manually designed
strategies, or the LLM-generated strategy introduced below. A list-level
Evaluator assigns
\begin{equation}
  u_j = E(L_j,x), \qquad
  j^\star=\arg\max_{j\in\mathcal{C}}u_j,\qquad
  L^\star=L_{j^\star},
  \label{eq:ge}
\end{equation}
where $\mathcal{C}$ is the active Generator pool. Unlike a pointwise ranker,
$E$ consumes the full list with user, item, position, and business context.
Its supervised heads cover point-level click and transaction behavior,
down-scroll behavior, list-level click outcomes, and effective cost per mille
(eCPM). Their calibrated
combination represents the deployment utility used to compare complete lists.

The Evaluator is intentionally separate from the LLM. It represents the
platform-side decision boundary, including user value, commercial value, and
experience constraints. The LLM policy searches for a personalized strategy
within this boundary.

\subsection{Strategy Generation Objective}

Let $\mathcal{A}$ be the schema-constrained strategy space and
$\pi_\theta(a\mid x)$ the LLM policy. The LLM produces one complete bundle
$a\in\mathcal{A}$ per generation. A deterministic tool process
$\toolprocess$ converts the semantic action into production parameters, and
the parameters control an isolated Generator $G_{\toolprocess(a)}$. The
learning objective is
\begin{equation}
  \max_\theta\;
  \mathbb{E}_{x\sim\mathcal{D},\,a\sim\pi_\theta(\cdot\mid x)}
  \left[E\!\left(G_{\toolprocess(a)}(I,x),x\right)\right],
  \label{eq:objective}
\end{equation}
subject to schema validity, execution constraints, and platform guardrails.
Because the bundle contains multiple dependent modules, the policy factorizes
autoregressively:
\begin{equation}
  \pi_\theta(a\mid x)
  =\prod_{m=1}^{M}\pi_\theta(a_m\mid x,a_{<m}).
  \label{eq:factorization}
\end{equation}
The small domain of each individual field does not make the problem
enumerable: the joint space grows exponentially, and later modules can
reinforce or override earlier decisions.

\section{\systemname}
\label{sec:method}

Figure~\ref{fig:overview} summarizes the execution and learning loop.
\systemname{} compiles request-conditioned strategies into one isolated
Generator, obtains list-level feedback from the atomic GE call, and closes the
loop through curriculum competition and routed OPD. The upstream intent model
supplies a current multi-intent state, while the strategy policy also observes
historical aggregates and raw recent behavior. This division avoids relearning
intent while retaining temporal evidence absent from a compressed state.

\subsection{Executable Strategy Bundle}

The policy emits one JSON object enclosed by a fixed answer delimiter. The
object contains five ordered executable modules spanning objective weights,
content-type and category preferences, experience constraints, and top-CTR
policies. The complete module definitions and value domains are detailed in
Appendix~\ref{app:action-space}.

Bounded, named fields make decisions reviewable, expose cross-module
interactions, and allow field-level limits. Sharing the schema across training,
replay, nearline generation, and serving reduces training-serving skew.

\subsection{Validation, Compilation, and Isolation}

An output passes through four deterministic gates. The parser first extracts
the answer block and parses JSON. Schema validation then checks module
completeness, ordering, field types, enumerations, and numeric ranges. The tool
process maps each semantic action to production parameters, including
Generator configuration, diversity rules, and top-position settings. Finally,
the compiled parameters are attached only to the designated LLM-controlled
Generator.

Invalid Student output receives a format penalty and skips Student tool
execution. In multi-Teacher training, a malformed Teacher is removed from the
routing set, while other valid candidates may still be evaluated. This
fail-closed behavior prevents missing fields or empty configurations from
becoming a shortcut. Configuration isolation also makes the replay reward
attributable to the LLM strategy rather than an accidental change to another
Generator.

\subsection{Production-Path Replay Environment}
\label{sec:replay}

Panel (c) of Figure~\ref{fig:overview} depicts our production-path replay
environment, which replaces a locally implemented user simulator. Each episode
is initialized from the input log of a real online request. The log preserves
the candidate set, its fine-ranking scores, and the request context required by
the re-ranking stage. Recall and fine-ranking inference are not replayed; the
episode starts from these fixed candidates and scores and executes the
production re-mixing and re-ranking path through isolated load-test traffic.
At replay time, the request resolves current online features and invokes the
current production models, including a fresh list-level Evaluator score for
every generated list.

Within one atomic GE invocation, the LLM-controlled Generator competes against
approximately ten production Generators, including perturbation-based and
context-aware re-estimation variants. All Generators operate on the same
request and candidate set, and all valid lists are scored by one Evaluator
invocation. A Generator that times out, fails execution, or returns an invalid
list is removed from the comparison; the remaining valid Generators are still
evaluated. Consequently, reward computation uses the valid Generator set of
that invocation.

Replay is isolated from user-facing serving. Its selected list is returned to
the training process but is never exposed to a user, and the request does not
write impression logs, attribution records, or business counters. An episode
is a single-step contextual decision: the environment does not synthesize
clicks or transactions and does not simulate a subsequent user-state
transition. The Evaluator output is therefore a proxy utility for the
resulting list rather than simulated or observed user feedback.

Repeated executions of the same logged request need not produce identical
results. Online features are resolved at replay time, and the Evaluator and
other production models may change with online releases. We therefore avoid
comparing absolute scores across replay invocations. Instead, the learning
signal is based on paired, within-request comparisons among Generators that
share one atomic GE invocation. This design preserves fidelity to the current
production re-ranking stack, while restricting optimization to the support of
the logged candidates and fixed fine-ranking scores. It is not an exact
reconstruction of the historical serving state or a long-horizon user
simulator.

\begin{figure*}[!t]
  \centering
  \includegraphics[width=0.8\textwidth]{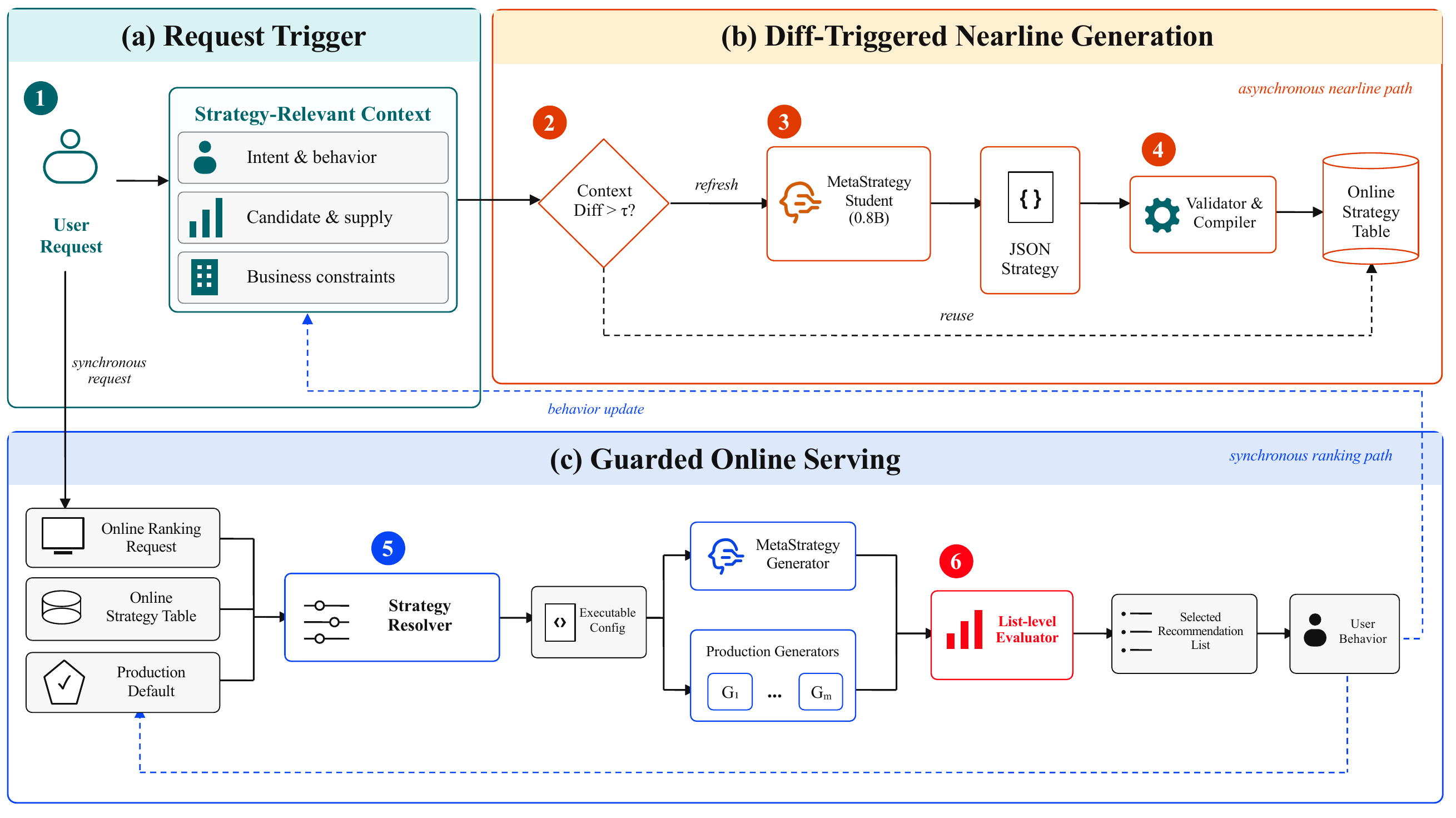}
  \caption{Nearline generation and guarded GE serving. Requests proceed to
  online ranking while a context-difference gate asynchronously reuses or
  refreshes the strategy. Valid Student output is compiled and published; the
  synchronous resolver selects a personalized or default configuration for the
  isolated \systemname{} Generator under the incumbent Evaluator.}
  \Description{A three-panel deployment pipeline. Panel a shows a user request
  entering the synchronous ranking path and forming strategy-relevant context
  from intent and behavior, candidate and supply information, and business
  constraints. Panel b compares the context with that of the current strategy.
  A refresh invokes the 0.8B MetaStrategy Student, which emits a JSON strategy
  that is validated, compiled, and written to an online strategy table; a reuse
  branch keeps the current entry. Panel c shows an online ranking request, the
  online strategy table, and the production default entering a strategy
  resolver. Its executable configuration instantiates the MetaStrategy
  Generator alongside production Generators. A list-level Evaluator selects
  the recommendation list, and subsequent user behavior updates future
  strategy context.}
  \label{fig:nearline-serving}
\end{figure*}

\subsection{Evaluator-Derived GE Reward}
\label{sec:reward}

For each replay invocation, the environment returns the scores of all valid
Generator lists and the Generator selected by the Evaluator. The
LLM-controlled Generator is compared with a designated production baseline
and the other valid production Generators under this shared evaluation state.

Let $u_S$ be the Student Generator score, $u_B$ the designated baseline score,
$N$ the number of valid competing lists, and
$\operatorname{rank}_S\in\{0,\ldots,N-1\}$ the descending rank of $u_S$. We
use three complementary rewards:
\begin{align}
  r_{\mathrm{sel}}
    &= \mathbb{I}[j^\star=S], \label{eq:rsel}\\
  r_{\mathrm{rank}}
    &= \frac{N-1-\operatorname{rank}_S}{\max(N-1,1)}, \label{eq:rrank}\\
  r_{\mathrm{lift}}
    &= \operatorname{clip}\!\left(
       \frac{u_S-u_B}{\max(|u_B|,\epsilon)},-c,c\right).
       \label{eq:rlift}
\end{align}
Selection is aligned with the final GE decision but sparse. Rank gives a dense
relative signal across all competitors. Lift measures continuous improvement
over the same-call production reference. The aggregate reward is
\begin{equation}
  r_{\mathrm{GE}}=
  w_{\mathrm{sel}}\cdot r_{\mathrm{sel}}+
  w_{\mathrm{rank}}\cdot r_{\mathrm{rank}}+
  w_{\mathrm{lift}}\cdot r_{\mathrm{lift}}.
  \label{eq:reward}
\end{equation}

For group-based policy optimization, multiple bundles are sampled for the
same request and the reward is normalized within the group:
\begin{equation}
  A^{\mathrm{RL}}_{q,k} =
  \frac{r_{q,k}-\mu_q}{\sigma_q+\epsilon}.
  \label{eq:grpo-adv}
\end{equation}
This advantage is used by a clipped policy objective. The reward depends on
the executed list, while the policy action remains the strategy bundle.

\subsection{Self-Competitive Strategy Curriculum}
\label{sec:curriculum}

The Evaluator is a learned proxy and may contain exploitable regularities. In
early training, we observe that a policy can repeatedly push bounded fields to
extreme values or emit nearly identical bundles for unrelated requests. Such a
policy may score well in replay while providing little request-level
adaptation.

Our curriculum turns the policy's dominant behavior into its next competitor,
as shown by the frequent-mode feedback path in Figure~\ref{fig:overview}(c).
At curriculum round $t$, the policy is trained against Generator pool
$\mathcal{C}_t$:
\begin{equation}
  \theta_t=\operatorname{RL\_Train}(\mathcal{D},\mathcal{C}_t).
\end{equation}
We execute valid sampled bundles and compute a canonical signature over the
compiled configuration,
$z=\operatorname{Sig}(\toolprocess(a))$. Using post-compilation signatures
merges JSON variants that are operationally equivalent. Let $\mathcal{Z}_t$
contain the $k$ most frequent signatures. The next pool is
\begin{equation}
  \mathcal{C}_{t+1} =
  \mathcal{C}_t \cup
  \{G(z):z\in\mathcal{Z}_t\}.
  \label{eq:curriculum}
\end{equation}
Each $G(z)$ is a frozen Generator that executes the corresponding strategy.
The next policy must outperform these now-explicit modes rather than obtain
reward by rediscovering them.

The curriculum is an outer loop across training jobs. It does not enumerate
the joint strategy space and does not require online execution of multiple LLM
outputs. Checkpoints from different rounds, random seeds, model scales, or
training preferences can form a complementary Teacher pool. We additionally
track compiled-signature diversity and bounded-action saturation; these are
diagnostics and optional regularizers, while Eq.~\ref{eq:curriculum} is the
mechanism that raises competition difficulty.

\subsection{Evaluator-Routed Reward-Augmented OPD}
\label{sec:opd}

The distillation branch in Figure~\ref{fig:overview}(c) compresses a pool of
expensive strategy policies into a compact Student. Routing is performed per
request, and a valid comparison requires all candidate strategies to see the
same replay state. For request $x$, the Student and $M$ Teachers generate
\begin{equation}
  a^S\sim\pi_\theta(\cdot\mid x),\qquad
  a^i\sim\pi_i(\cdot\mid x),\quad i=1,\ldots,M.
\end{equation}
Every valid bundle is compiled into an isolated Generator rule. The Student,
Teachers, and production baselines are submitted in one replay call, so their
lists share the original request, candidates, and Evaluator state.

Teacher routing excludes the Student. Among Teachers that pass parsing,
schema, tool execution, and Evaluator scoring, we choose
\begin{equation}
  i^\star(x)=
  \arg\max_{i\in\mathcal{V}(x)}
  E\!\left(G_{\toolprocess(a^i)}(I,x),x\right),
  \label{eq:routing}
\end{equation}
where $\mathcal{V}(x)$ is the valid Teacher set. A deterministic identifier
breaks ties. If $\mathcal{V}(x)$ is empty, the request is routed to
\texttt{\_\_none\_\_}: its RL reward is retained and the distillation term is
masked.

OPD is evaluated on the Student's sampled token trajectory, not on a
Teacher-forced target sequence. At response token $\ell$,
\begin{equation}
  d_\ell(x)=
  \log\pi_\theta(a^S_\ell\mid x,a^S_{<\ell})
  -\log\pi_{i^\star(x)}(a^S_\ell\mid x,a^S_{<\ell}).
  \label{eq:opd-ratio}
\end{equation}
Under Student sampling, this log-ratio estimates the reverse direction of the
Kullback--Leibler (KL) divergence,
$\mathrm{KL}(\pi_\theta\|\pi_{i^\star})$. We combine it directly with the
reward advantage:
\begin{equation}
  A^{\mathrm{mix}}_{\ell} =
  A^{\mathrm{RL}}-\lambda_{i^\star(x)}d_\ell(x),
  \label{eq:mixed-advantage}
\end{equation}
and use $A^{\mathrm{mix}}_\ell$ in the standard clipped policy objective. Only
the routed Teacher contributes reference log-probabilities. This is not an
additional Teacher-forced cross-entropy loss or a standalone KL loss.

Keeping $A^{\mathrm{RL}}$ matters. Pure imitation cannot exceed the routed
Teacher and can propagate poor supervision when every Teacher is weak. The
mixed objective preserves exploration against the production GE reward, uses
the best comparable Teacher when one is available, and degrades to RL-only
training when routing is unavailable.

\subsection{Two-Stage Training Procedure}

Teacher construction iteratively trains policies against curriculum pools,
adds frequent compiled modes as competitors, and retains complementary
checkpoints. Student compression atomically evaluates one bundle from each
policy, routes the best valid Teacher, and updates the Student with
Eq.~\ref{eq:mixed-advantage}. The Teacher pool is offline-only; each nearline
refresh invokes only the deployed Student.

\section{Production Deployment}
\label{sec:deployment}

\systemname{} is deployed in the Taobao homepage \guly{} feed, where products,
content posts, short videos, live streams, and sponsored items compete under
user, transaction, commercial, and experience objectives. As
Figure~\ref{fig:nearline-serving} shows, LLM inference runs in an asynchronous
nearline branch; the synchronous path performs only strategy lookup and
deterministic resolution. This temporal and structural decoupling adds no
observable RT overhead and leaves incumbent Generators, the Evaluator, and fallback
behavior unchanged.

\textbf{Diff-triggered generation.}
As shown in Figure~\ref{fig:nearline-serving}, a request proceeds directly to
the synchronous ranker while a nearline branch assembles strategy-relevant
context from intent and behavior, candidate and supply summaries, and business
constraints. A feature-aware difference gate invokes the deployed 0.8B Student
only when no valid entry exists or the context has changed sufficiently;
otherwise it reuses the latest strategy. The Student emits the same typed JSON
bundle used during training. Validation and compilation precede publication to
the online strategy table, so failed refreshes cannot replace the latest valid
entry.

\textbf{Guarded GE serving.}
The synchronous path performs no LLM inference. A deterministic resolver reads
the latest valid entry or the production default and attaches the executable
configuration only to the isolated \systemname{} Generator. Its list competes
with unchanged production Generators under the incumbent Evaluator and is
exposed only when selected by the atomic GE call. Temporal decoupling,
validation-before-publication, deterministic fallback, and structural
isolation preserve online RT, support immediate rollback, and introduce no
observable latency increase. Appendix~\ref{app:deployment} gives the complete
refresh and fallback protocol.

\begin{table*}[!t]
  \centering
  \caption{Production-path replay on Taobao Homepage \guly{}. Pointwise and
  Evaluator columns are relative changes from BaseG. Production rows use the
  incumbent GE call; each LLM Generator is added individually to the same
  pinned incumbent pool. All rows use the same request set and are interleaved
  within one replay window. Best and second-best comparable values are bolded
  and underlined, respectively. ``--'' denotes a metric that is not applicable.}
  \label{tab:offline-main}
  \resizebox{\textwidth}{!}{%
  \begin{tabular}{lccccccccc}
    \toprule
    Method & $\Delta$pCTR $\uparrow$ & $\Delta$pCVR $\uparrow$ &
    $\Delta$pIPV $\uparrow$ & $\Delta$pGMV $\uparrow$ &
    $\Delta$Evaluator $\uparrow$ & Valid $\uparrow$ & Selected $\uparrow$ &
    $\Delta$GE $\uparrow$ & Beat Base $\uparrow$ \\
    \midrule
    \multicolumn{10}{l}{\textit{Production ranking Generators}} \\
    Base G  &  -- &  -- &  -- &  -- &  -- &  -- &  6.97\% &  -- & -- \\
    GNR G  & -2.43\% & +2.08\% & -2.75\% & +14.04\% & +1.68\% & -- & 6.09\% & -- & 37.4\% \\
    NAR (non-autoregressive) G  & \underline{+1.64\%} & +15.24\% & \underline{+6.46\%} & +17.03\% & \textbf{+2.75\%} & -- & \underline{13.7\%} & -- & \textbf{63.4\%} \\
    Single-objective perturbation G  & \textbf{+13.82\%} & -4.01\% & \textbf{+11.81\%} & +1.05\% & +1.21\% & -- & 4.94\% & -- & 50.1\% \\
    \midrule
    \multicolumn{10}{l}{\textit{LLM strategy Generators}} \\
    Prompt-only (0.8B) & -2.82\% & +8.17\% & +0.13\% & +12.65\% & -0.29\% & 4.39\% & 0.39\% & +0.02\% & 1.8\% \\
    Prompt-only (4B) & -3.41\% & +13.67\% & -0.41\% & +13.81\% & +0.42\% & 80.86\% & 7.87\% & +0.34\% & 35.69\% \\
    RL (0.8B) & -0.48\% & +10.10\% & +2.51\% & +13.58\% & +1.05\% & 97.12\% & 8.56\% & +0.38\% & 46.88\% \\
    RL (4B) & -1.07\% & +12.54\% & +0.96\% & +14.23\% & +1.58\% & \underline{98.85\%} & 7.93\% & +0.45\% & 45.86\% \\
    Curriculum RL (4B) & -5.70\% & \underline{+18.47\%} & -2.95\% & \underline{+18.88\%} & +0.56\% & \textbf{98.89\%} & 11.27\% & \underline{+0.57\%} & 45.65\% \\
    Routed reward-augmented OPD (0.8B) & -0.68\% & \textbf{+22.69\%} & +1.65\% & \textbf{+37.76\%} & \underline{+2.53\%} & 98.03\% & \textbf{16.24\%} & \textbf{+0.73\%} & \underline{62.60\%} \\
    \bottomrule
  \end{tabular}%
  }
\end{table*}
\section{Experiments}
\label{sec:experiments}

The evaluation addresses four questions:
\begin{description}
  \item[RQ1:] Does executable strategy generation improve list quality and
  add incremental value beyond incumbent production Generators?
  \item[RQ2:] How do model scale and successive training stages affect
  executability, strategy concentration, and contribution to the GE pool?
  \item[RQ3:] Can routed reward-augmented OPD compress the 4B policies into a
  competitive 0.8B Student?
  \item[RQ4:] What user and platform outcomes does the method produce under
  online deployment?
\end{description}

\subsection{Experimental Setup}

\textbf{Industrial replay dataset.} We collect online request logs from Taobao
Homepage \guly{} over eight consecutive days, randomly sampling 8,192 requests
per day for 65,536 requests in total. Days 1--6 form the training set (49,152
requests), Day 7 the validation set (8,192), and Day 8 the test set (8,192).
Products, content posts, short videos, live streams, and sponsored items share
this recommendation surface. Each log preserves the strategy-policy context,
candidate set, fine-ranking scores, and constraints required by the re-ranking
stage. During replay, these candidates and scores remain fixed, while current
online features, production Generators, and the list-level Evaluator execute
through the isolated production path. Test-period feedback is excluded from RL,
curriculum construction, OPD, prompt selection, and hyperparameter tuning.

\textbf{Replay protocol.} The production rows are measured in the
unmodified incumbent Generator pool $\mathcal{P}$, and each LLM benchmark
Generator $G_j$ is added individually to the same pool. All methods use exactly
the same logged requests and pinned production-model and Evaluator versions.
Their calls are interleaved within the same replay window to balance temporal
variation in online feature lookups. Within an atomic call, all Generators share
the logged request, candidate set, fixed fine-ranking scores, online feature
state, and one Evaluator invocation. We compute utility relative to BaseG or
the incumbent pool within that invocation. A strategy is valid when it passes
parsing and compilation and returns an executable list. Only valid lists enter
the GE comparison; validity is measured over all test requests, and a request
without a valid benchmark list contributes zero to that method's selection
rate and incremental GE lift.

\textbf{Production Generator baselines.}
\begin{itemize}
  \item \textbf{Base G} greedily fills each position with the remaining card
  having the highest production rank score while satisfying the same hard
  filtering, position, density, and diversity constraints used by all methods.
  It is the reference method for relative improvements.
  \item \textbf{GNR G} \cite{meng2025generative} is a sequential, context-aware production Generator.
  At each position, it conditions on the cards already placed, re-estimates
  click propensity for the remaining candidates, and selects the next card
  under the production constraints.
  \item \textbf{NAR G} \cite{xu2026omgrec} predicts position-specific scores in
  parallel and uses a non-autoregressive list constructor. It captures
  position preference without sequential inference.
  \item \textbf{Single-objective perturbation G} starts from the BaseG scoring
  formula and perturbs one business objective while retaining the same rule
  stack. We instantiate the production perturbation configurations before
  evaluating on the test period; individual target variants are reported
  separately when multiple configurations are used.
\end{itemize}

\textbf{LLM Generator variants.}
\begin{itemize}
  \item \textbf{Prompt-only} applies the common prompt, JSON action schema,
  validation, compiler, and execution engine without parameter updates from
  replay feedback. We evaluate 0.8B and 4B policies.
  \item \textbf{RL} optimizes the three-component reward in
  Eq.~\ref{eq:reward} against the original production pool. We train both 0.8B
  and 4B policies without feeding frequent strategies back as competitors.
  \item \textbf{Curriculum RL} applies the self-competitive strategy curriculum
  to the 4B policy, progressively adding frequent compiled strategies to the
  competing Generator pool.
  \item \textbf{Routed reward-augmented OPD} distills two 4B checkpoints into
  the deployable 0.8B Student. For each request, the Student learns from the
  valid Teacher selected by the shared Evaluator call while retaining its own
  GE reward.
\end{itemize}

\subsection{Formal Evaluation Metrics}
\label{sec:metrics}

Let $L_q^j$ be the length-$K_q$ list produced by Generator $G_j$ on request
$q$, and let $\mathcal{V}_j$ be the requests on which it returns a valid list.
For objective $m\in\{\mathrm{CTR},\mathrm{CVR},\mathrm{IPV},\mathrm{GMV}\}$,
the pointwise list diagnostic averages the request-preserved fine-ranking score
$s_m$ first within each list and then over valid requests:
\begin{equation}
  S_m(G_j)=\frac{1}{|\mathcal{V}_j|}
  \sum_{q\in\mathcal{V}_j}\frac{1}{K_q}
  \sum_{i\in L_q^j}s_m(i,x_q).
  \label{eq:pointwise-list-metric}
\end{equation}

Let $\mathcal{C}_{q,j}$ be the valid Generator set in the atomic call for row
$j$, and let $\mathcal{P}$ be the incumbent pool. Using the production
tie-breaking rule, define
\begin{equation}
  j_q^\star=\operatorname{TieBreakArgMax}_{g\in\mathcal{C}_{q,j}}
  E(L_q^g,x_q).
  \label{eq:actual-ge-winner}
\end{equation}
Within the same call, define
\begin{equation}
  u_{q,j}=E(L_q^j,x_q),\quad
  u_q^{\mathcal{P}}=\max_{g\in\mathcal{P}}E(L_q^g,x_q).
\end{equation}
With $u_{q,B}$ denoting the BaseG score from the same call, Evaluator lift is
\begin{equation}
  \Delta_{\mathrm{Eval}}(G_j)=
  \frac{1}{|\mathcal{V}_j|}\sum_{q\in\mathcal{V}_j}
  \frac{u_{q,j}-u_{q,B}}{\max(|u_{q,B}|,\epsilon)}.
  \label{eq:eval-lift}
\end{equation}
We report validity $|\mathcal{V}_j|/N$ and the selection rate
\begin{equation}
  \mathrm{SR}(G_j)=\frac{1}{N}\sum_{q=1}^{N}
  \mathbb{I}[q\in\mathcal{V}_j]\mathbb{I}[j_q^\star=j].
  \label{eq:selected-rate}
\end{equation}
The beat-Base rate replaces the incumbent reference with $u_{q,B}$. For the
primary offline endpoint, set $\widetilde u_{q,j}=-\infty$ for invalid outputs
and $\widetilde u_{q,j}=u_{q,j}$ otherwise. Incremental GE lift is
\begin{equation}
  \Delta_{\mathrm{GE}}(G_j)=
  \frac{1}{N}\sum_{q=1}^{N}
  \frac{\max(u_q^{\mathcal{P}},\widetilde u_{q,j})-u_q^{\mathcal{P}}}
  {\max(|u_q^{\mathcal{P}}|,\epsilon)}.
  \label{eq:ge-incremental-lift}
\end{equation}
Finally, for any strategy module $h$, let $z^{(h)}_{q,j}$ be its compiled
decision from $G_j$ on request $q$. Its concentration is
\begin{equation}
  F_{\mathrm{top1}}(G_j;h)=
  \max_z\frac{1}{|\mathcal{V}_j|}
  \sum_{q\in\mathcal{V}_j}\mathbb{I}[z^{(h)}_{q,j}=z].
  \label{eq:strategy-top1}
\end{equation}
Lower values indicate broader use of module $h$'s action space. All
comparisons are paired by request; an invalid output contributes zero to
selection and incremental lift.

\subsection{Offline Evaluation}

Table~\ref{tab:offline-main} answers RQ1--RQ3. We prioritize
$\Delta_{\mathrm{GE}}$, the incremental utility of adding one Generator to
the incumbent pool. Evaluator lift and beat-Base measure standalone list
quality, selection measures competition within an atomic call, and the
pointwise columns diagnose which objectives move.

\textbf{Production baselines.}
NAR is the strongest incumbent, with $+2.75\%$ Evaluator lift, $13.70\%$
selection, and $63.40\%$ beat-Base. GNR obtains $+1.68\%$ Evaluator lift but is
selected on only $6.09\%$ of calls and beats BaseG on $37.40\%$ of requests.
The single-objective perturbation produces the largest pCTR and pIPV gains
($13.82\%$ and $11.81\%$), yet yields only $+1.21\%$ Evaluator lift and
$4.94\%$ selection. These contrasts show that improving an isolated predictor
does not ensure a better heterogeneous list under the platform objective.

\textbf{Scale and replay RL.}
Scaling the prompt-only policy from 0.8B to 4B raises validity from $4.39\%$
to $80.86\%$, selection from $0.39\%$ to $7.87\%$, and
$\Delta_{\mathrm{GE}}$ from $0.02\%$ to $0.34\%$. Thus, scale primarily
improves the ability to emit executable strategies. Replay RL makes the 0.8B
policy $97.12\%$ valid and raises its selection and $\Delta_{\mathrm{GE}}$ to
$8.56\%$ and $0.38\%$, slightly exceeding the prompt-only 4B policy on both
GE metrics. At 4B, RL further increases Evaluator lift to $1.58\%$ and
$\Delta_{\mathrm{GE}}$ to $0.45\%$, although selection remains $7.93\%$.
Model scale therefore helps feasibility, whereas executed-list reward supplies
the alignment needed for GE contribution.

\textbf{Curriculum and routed OPD.}
Relative to direct RL at matched 4B scale, Curriculum RL keeps validity unchanged
($98.85\%$ versus $98.89\%$) while raising selection from $7.93\%$ to
$11.27\%$ and $\Delta_{\mathrm{GE}}$ from $0.45\%$ to $0.57\%$. Its pCTR and
pIPV decrease while pCVR and pGMV increase, illustrating that the curriculum
optimizes joint list utility rather than simultaneous pointwise dominance.

The routed 0.8B Student is $98.03\%$ valid and obtains the highest selection
($16.24\%$) and incremental lift ($0.73\%$), while ranking second in Evaluator
lift ($2.53\%$) and beat-Base ($62.60\%$). It exceeds both 4B training variants
on the two GE contribution metrics despite its smaller serving size. NAR has
slightly higher standalone Evaluator lift, but the Student's leading
$\Delta_{\mathrm{GE}}$ shows greater marginal complementarity to the incumbent
pool. Together, the results show that executable strategy generation adds
incremental list value (RQ1), scale and replay learning play distinct roles
(RQ2), and routed OPD transfers complementary Teacher behavior into the
deployable Student (RQ3).

\subsection{Ablation of Strategy-Collapse Mitigation}

\begin{figure}[t]
  \centering
  \includegraphics[width=\columnwidth]{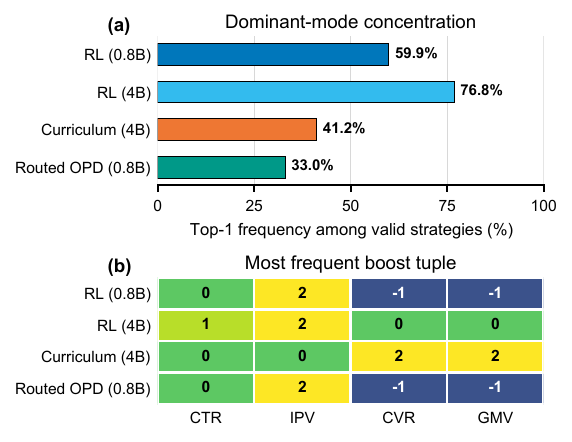}
  \Description{A collapse-mitigation ablation using the
  ranking-weight-boost strategy. Panel a is a horizontal bar chart of the most
  frequent tuple: RL 0.8B, 59.9 percent; RL 4B, 76.8 percent; Curriculum RL
  4B, 41.2 percent; and Routed OPD 0.8B, 33.0 percent. Panel b shows the
  corresponding CTR, IPV, CVR, and GMV tuples: 0, 2, -1, -1; 1, 2, 0, 0;
  0, 0, 2, 2; and 0, 2, -1, -1.}
  \caption{Collapse mitigation for \texttt{ranking\_weight\_boost}:
  (a) Top-1 complete-tuple frequency and (b) dominant tuple in
  $(\mathrm{CTR},\mathrm{IPV},\mathrm{CVR},\mathrm{GMV})$ order.}
  \label{fig:strategy-concentration}
  \vspace{-1.2\baselineskip}
\end{figure}

We use \texttt{ranking\_weight\_boost} as a representative strategy. It assigns
one value in $\{-2,-1,0,1,2\}$ to each of CTR, IPV, CVR, and GMV, giving 625
complete tuples. Figure~\ref{fig:strategy-concentration} applies
Eq.~\ref{eq:strategy-top1} to measure concentration among valid outputs. We
interpret it with validity and GE utility in Table~\ref{tab:offline-main}, so a
lower frequency cannot be attributed to failed execution.

\begin{table}[!t]
  \centering
  \caption{Seven-day online A/B test. Bold values are statistically
  significant.}
  \label{tab:online}
  \scriptsize

  \begin{tabular}{@{}cccccc@{}}
    \toprule
    \shortstack{Exposure PV $\uparrow$} &
    \shortstack{Click PV $\uparrow$} & IPV $\uparrow$ &
    \shortstack{Transaction\\Count $\uparrow$} &
    \shortstack{Transaction\\Amount $\uparrow$} &
    \shortstack{Ad Cost $\uparrow$} \\
    \midrule
    \textbf{+1.49\%} & \textbf{+2.11\%} & \textbf{+3.12\%} &
    $-0.24$\% & \textbf{+2.83\%} & +0.87\% \\
    \bottomrule
  \end{tabular}
  \vspace{-0.8\baselineskip}
\end{table}

\textbf{Scale control.}
Direct RL concentrates $59.9\%$ of 0.8B outputs on $(0,2,-1,-1)$ and $76.8\%$
of 4B outputs on $(1,2,0,0)$. Thus, added capacity alone does not prevent
collapse under the same Evaluator-derived reward.

\textbf{Curriculum ablation.}
At 4B, the self-competitive curriculum reduces Top-1 frequency from $76.8\%$
to $41.2\%$ while validity remains unchanged ($98.85\%$ versus $98.89\%$).
Selection rises from $7.93\%$ to $11.27\%$ and $\Delta_{\mathrm{GE}}$ from
$0.45\%$ to $0.57\%$, indicating broader request-conditioned choices without
sacrificing GE contribution.

\textbf{Routed-OPD ablation.}
At 0.8B, routed reward-augmented OPD lowers Top-1 frequency from $59.9\%$ to
$33.0\%$. Relative to direct RL, validity rises from $97.12\%$ to $98.03\%$,
selection from $8.56\%$ to $16.24\%$, and $\Delta_{\mathrm{GE}}$ from $0.38\%$
to $0.73\%$. The matched-scale ablations therefore show that curriculum
competition and routed OPD both mitigate collapse while improving list-level
utility.

\subsection{Online A/B Test}

We conduct a seven-day online A/B test in July 2026. Users are assigned to stable control and treatment buckets via a deterministic user-level hash. Each bucket represents 1\% of traffic from a large-scale production service and contains millions of users. A preceding A/A test
finds no material imbalance. Outcomes are normalized per user. Ad cost denotes
advertising spend and is therefore a positive platform commercial-value metric.

The control retains the incumbent GE pool, while the treatment adds the final
routed reward-augmented OPD 0.8B Generator with all other ranking components
and eligibility rules fixed. \systemname{} wins 27.93\% of treatment-side
atomic GE calls. Table~\ref{tab:online} shows gains of 1.49\% in exposure PV,
2.11\% in click PV, 3.12\% in IPV, 2.83\% in transaction amount, and 0.87\% in
ad cost; transaction count changes by $-0.24$\%. The platform's standard
procedure finds the gains in exposure PV, click PV, IPV, and transaction amount
significant. Transaction count and ad cost are not statistically significant,
although the ad-cost direction is positive for platform commercial value.
Deployed on this recommendation surface, \systemname{} serves hundreds of
millions of users daily, demonstrating its effectiveness in real-world production.

\section{Conclusion}

\systemname{} reframes generative ranking as request-conditioned,
schema-bounded executable strategy generation. Its typed policy and compiler instantiate an isolated
Generator that competes under the incumbent Evaluator. Production-path replay,
self-competitive curriculum, and Evaluator-routed reward-augmented OPD yield a
deployable 0.8B Student with $+0.73\%$ incremental GE lift offline. In a
seven-day online A/B test, click PV, IPV, and transaction amount increase by
$2.11\%$, $3.12\%$, and $2.83\%$, respectively, with no added ranking RT.
These results establish a practical LLM strategy layer for personalized
ranking. Future work will study robustness to Evaluator updates, broader typed
action spaces, and long-term user feedback beyond fixed-candidate replay.

\clearpage
\bibliographystyle{ACM-Reference-Format}
\bibliography{references}

\clearpage
\appendix

\section{Complete Strategy Action Schema}
\label{app:action-space}

The action schema is the contract between the language policy and the
production re-ranking system. It intentionally excludes item identifiers,
free-form code, and direct permutations. Instead, the policy must emit a
complete bundle of typed, bounded decisions whose semantics are stable across
offline training, production-path replay, nearline generation, and online
execution. The ordinal values in Table~\ref{tab:action-space} express relative
preferences or adjustment strengths; they do not replace the incumbent
fine-ranking scores or bypass hard business and experience constraints.

The five modules expose complementary controls. Weight adjustments alter the
balance among engagement and transaction targets;
card-type and category preferences control supply composition;
experience constraints regulate exposure, purchase, and density patterns; and
top-CTR switches activate audited page-specific templates. Modules are
interpreted jointly because a valid strategy may combine several adjustments
for one request. Every output must include the required modules in the
prescribed order and satisfy field types, enumerations, and value ranges.
Unknown, missing, or out-of-range fields cause the bundle to be rejected.

After validation, a versioned compiler maps the semantic actions to executable
production parameters. The downstream resolver then applies eligibility checks
and hard constraints before list construction. This separation keeps the LLM
action auditable and permits the concrete fields or bounds to evolve through a
matched schema-and-compiler update without changing the learning formulation.
Table~\ref{tab:action-space} reports the schema used in the current deployment.

\begin{table}[H]
  \centering
  \caption{Structured strategy space in the current deployment. The method
  requires typed executable modules; the particular fields are configurable.}
  \label{tab:action-space}
  \small
  \setlength{\tabcolsep}{3pt}
  \begin{tabular}{@{}p{0.34\columnwidth}p{0.59\columnwidth}@{}}
    \toprule
    Module & Decision and domain \\
    \midrule
    \texttt{ranking\_weight\_}\allowbreak\texttt{boost} &
    CTR, IPV, CVR, and GMV adjustments;
    $\{-2,-1,0,1,2\}$ per field. \\
    \texttt{cardtype\_}\allowbreak\texttt{preference} &
    Preferences for auction, advertising, promotion, video, live, and
    professional content; $\{-2,-1,0,1,2\}$ per field. \\
    \texttt{category\_}\allowbreak\texttt{preference} &
    Preference strength for eligible categories;
    $\{-2,-1,0,1,2\}$ per key. \\
    \texttt{experience\_}\allowbreak\texttt{constraints} &
    Exposure, purchase, and density constraints;
    $\{-2,-1,0,1,2\}$ per field. \\
    \texttt{top\_ctr\_}\allowbreak\texttt{strategy} &
    Page-specific top-CTR and reverse-order switches;
    $\{0,1\}$ per page group. \\
    \bottomrule
  \end{tabular}
\end{table}

\section{Detailed Production Deployment}
\label{app:deployment}

\subsection{Request Trigger and Context Fork}

Every ranking request creates two logically decoupled paths. The synchronous
branch proceeds directly to the online ranker. In parallel, the nearline branch
assembles a strategy-relevant context $h_t$ from intent and behavior, candidate
and supply summaries, and active business constraints. The behavior signals
combine the upstream intent model's real-time output, historical user
statistics, and the ordered sequence of recent interactions. The synchronous
branch uses the latest available entry, while a refreshed strategy becomes
visible only after publication to the online table.

\subsection{Diff-Triggered Generation}

Let $\bar h_t$ denote the context associated with the latest valid table entry.
The nearline service invokes the deployed Student when no valid entry exists or
when
\begin{equation}
  \Delta(h_t,\bar h_t)>\tau,
  \label{eq:refresh-trigger}
\end{equation}
where $\Delta$ measures feature-aware changes in strategy inputs. Otherwise,
the service reuses the current entry. On refresh, the validator and compiler
enforce schema and field-level bounds before translating the JSON bundle into
executable ranking parameters. Only a valid configuration is published. A
failed refresh does not replace the latest valid entry; when no usable entry is
available, synchronous serving selects the production default.

\subsection{Strategy Resolution and Isolation}

For each ranking request, a deterministic resolver reads the latest table entry
and the production default, checks validity, and returns an executable
configuration. A valid personalized entry takes precedence; an invalid,
missing, or expired entry selects the default. The resolved configuration is
attached only to the \systemname{} Generator. No incumbent Generator,
Evaluator, or fallback branch is modified, and the \systemname{} list is
exposed only when it wins the atomic GE comparison. The system therefore
supports gradual traffic ramp-up, request-level attribution, and immediate
rollback without on-request model updates.

\section{Limitations and Responsible Deployment}
\label{sec:limitations}

\systemname{} optimizes a learned Evaluator rather than online value directly.
Self-competition raises the cost of exploiting frequent strategy modes, but it
cannot remove Evaluator misspecification. The Evaluator and other production
models can also change across online releases. Atomic replay makes Generator
scores comparable within a request, not across replay invocations or model
versions.

Replay fixes the logged candidate set and its fine-ranking scores, then
executes the current production re-ranking path. It therefore cannot evaluate
gains that require different retrieval candidates or upstream fine-ranking
predictions. Nor does it simulate clicks, transactions, or later user-state
transitions. Replay results measure list-level proxy utility within the support
of the logged request and must be validated by randomized online experiments.

The typed action schema further limits the policy to mechanisms exposed by the
production ranker. This restriction improves auditability and rollback, but it
may exclude useful strategies. Generated configurations can alter commercial
exposure and content composition, so deployment requires field-level bounds,
fallback behavior, traffic ramp-up, segment analysis, and continuous user
experience monitoring.

\end{document}